\documentclass[authorversion]{sigchi-ext}
\usepackage{booktabs} % For formal tables
\usepackage{ccicons}  % For Creative Commons citation icons
\usepackage[T1]{fontenc}

\let\autocite\cite

\ccsdesc[500]{Human-centered computing~Empirical studies in collaborative and social computing}
\ccsdesc[300]{Security and privacy~Social aspects of security and privacy}
\ccsdesc[300]{Human-centered computing~Empirical studies in ubiquitous and mobile computing}

\usepackage{booktabs}
\usepackage{longtable}
\usepackage{array}
\usepackage{multirow}
\usepackage{wrapfig}
\usepackage{float}
\usepackage{colortbl}
\usepackage{pdflscape}
\usepackage{tabu}
\usepackage{threeparttable}
\usepackage{threeparttablex}
\usepackage[normalem]{ulem}
\usepackage{makecell}
\usepackage{xcolor}

\usepackage{graphicx}
\usepackage{dcolumn}
\usepackage{enumitem}
\usepackage{calc}
\usepackage{textpos}

\DeclareCaptionType{sidebar}[Sidebar][List of Sidebars]

\author{%
    \alignauthor{%
    \begin{tabbing}
    \textbf{Martin J. Kraemer} \quad\quad\quad\quad\quad\quad\= \textbf{Ulrik Lyngs}\\
            \affaddr{University of Oxford} \> \affaddr{University of Oxford} \\
    \affaddr{United Kingdom} \> \affaddr{United Kingdom} \\
        \email{martin.kraemer@cs.ox.ac.uk} \>   \email{ulrik.lyngs@cs.ox.ac.uk} \\
        \\
    \textbf{Helena Webb} \> \textbf{Ivan Flechais}\\
            \affaddr{University of Oxford} \> \affaddr{University of Oxford} \\
    \affaddr{United Kingdom} \> \affaddr{United Kingdom} \\
       \email{helena.webb@cs.ox.ac.uk} \>   \email{ivan.flechais@cs.ox.ac.uk}
    \end{tabbing}
     } 
  }

\copyrightinfo{\scriptsize Permission to make digital or hard copies of part or all of this work for personal or classroom use is granted without fee provided that copies are not made or distributed for profit or commercial advantage and that copies bear this notice and the full citation on the first page. Copyrights for third-party components of this work must be honored. For all other uses, contact the owner/author(s). \\
{\emph{CHI '20 Extended Abstracts, April 25--30, 2020, Honolulu, HI, USA.}} \\
\copyright~2020 Copyright is held by the author/owner(s). \\
ACM ISBN 978-1-4503-6819-3/20/04. \\
DOI: https://doi.org/10.1145/3334480.3382972 }
\begin{document}

\title{Further Exploring Communal Technology Use in Smart Homes: Social Expectations}

\maketitle

\begin{abstract}
  Device use in smart homes is becoming increasingly communal, requiring cohabitants to navigate a complex social and technological context. In this paper, we report findings from an exploratory survey grounded in our prior work on communal technology use in the home \autocite{Kraemer2019}. The findings highlight the importance of considering qualities of social relationships and technology in understanding expectations and intentions of communal technology use. We propose a design perspective of social expectations, and we suggest existing designs can be expanded using already available information such as location, and considering additional information, such as levels of trust and reliability.
\end{abstract}

\begin{textblock}{4.9}[0,0](-5.64,-1.5)  
\keywords{communal technology use, smart home, multi-user}
\printccsdesc
\end{textblock}

\hypertarget{introduction}{%
\section*{Introduction}\label{introduction}}
\addcontentsline{toc}{section}{Introduction}

Internet-connected devices in our homes such as smart lights or voice assistants are becoming increasingly ubiquitous, easy to use, and unobtrusive while their use may be shared by many residents within the same household. 
Because households share not only resources but also responsibilities, not everyone in a household participates equally in setting up, managing, and using internet-connected devices.
This can lead to challenges such as a \emph{power imbalance} due to the accumulation of skill, competence, and knowledge for everyday use among more apt users (technology drivers \cite{Geeng2019}).
Researchers and designers have proposed and implemented different ways of interaction, kinds of controls, and features (e.g.~multiple accounts) to counter such imbalances.
However, how these can be put into practice for the benefit of the whole household remains a question of ongoing research \autocite{Jakobi2018,Geeng2019,Kraemer2019,Garg2019,Yao2019}.

\marginpar{%
\fbox{%
\footnotesize
\begin{minipage}{0.975\marginparwidth}
\textbf{Survey Outline}\\
\textit{{Additional Demographics}}
\begin{enumerate}[leftmargin=20pt,label=(\roman*)]
\item Attitude -- how much do you agree with each of the following statements?
\item Please introduce your own household using nicknames for each person living with you
\end{enumerate}  
\textit{{Part (A): Roles and responsibilities}}
\begin{enumerate}[leftmargin=20pt,label=(\roman*),topsep=5pt]
\item  (a) Who would configure the system?; (b) Are there other household members who could configure the system?; (c) How is the system going to be used? 
\begin{flushleft}
\scriptsize \textbf{vignettes}: (1) system: security, lights, thermostat
\end{flushleft}

\end{enumerate}  
\vspace{2pt}\hspace{20pt}\parbox{\linewidth-20pt}{\color{gray} questions (ii) + (iii) of part (A) are \textbf{not} part of this paper}\vspace{2pt}
\begin{enumerate}[leftmargin=20pt,label=(\roman*)]
\setcounter{enumi}{3}
\item Adjusting smart home system -- How would you recommend Grace and Oliver accommodate for their guests? \begin{flushleft} \scriptsize \textbf{vignettes}: (1) duration: weekend, week;
(2) social: niece/nephew, mum/dad, close friends, colleagues \end{flushleft}
\item Responsibility toward cohabitants -- How important is it for them to make sure that adults in the house ...  \begin{flushleft} \scriptsize \flushleft \textbf{vignettes}: (1) devices: security, light, thermostat, voice assistant, television
(2) social: partner, housemates, parents \end{flushleft}
\end{enumerate} 
{\color{gray}
\textit{Parts (B) + (C)} are \textbf{not} part of this paper}
\captionof{sidebar}{Outline includes additional demographics and details on parts (A-C) where relevant to the paper}\label{sidebar:survey-outline}
\end{minipage}}}

Communal use is socially constructed within the context of the home, and research is needed to enable all members ``to take an active part in controlling set up, evolution and destruction'' \autocite{Rogers2006}. We previously interviewed 36 participants to understand \emph{communal use} of internet-connected technology in the home \autocite{Kraemer2019}. We reported individual and communal considerations pertaining to people and technology in four themes: (1) dealing with technology, (2) sharing personal devices, (3) using shared devices, and (4) dealing with guests and visitors.
The findings illustrate that householders consider---not only their own but also their cohabitants'---attitude, preferences, skills, and competencies in arranging and accommodating for situations of communal use; the relevance of these aspects in practice, however, warrants further investigation.

To explore when and how they are considered, we designed a survey using scenarios and themes from our qualitative data. The survey explored how qualities of different systems and different social groups impact considerations of internet-connected technology use in the home \autocite{Kraemer2019}. We presented respondents with several situations of device use in communal settings to collect expressions of their own preferences, experiences, and sense-making.

This paper focuses on navigating communal use between permanent household members and their potential guests. Our findings show how well-intentioned efforts to be a responsible and respectful household member and technology owner meet complex social relationships that require thoughtful consideration. We suggest a perspective of social expectation to inform future design and highlight how already available information, such as location, and additional information, such as types of relationships or levels of trust, can provide useful contextual information for designers and users alike.

\hypertarget{methodology}{%
\section{Methodology}\label{methodology}}

From themes `using shared devices' and `guests and visitors'~\autocite{Kraemer2019}, we derived 14 questions (details available online: \url{doi.org/10.17605/OSF.IO/TZBVQ}). Three HCI researchers from our department commented on possible biases and the order of questions, and one non-expert user completed a cognitive interview \autocite{willis2004cognitive}. In May 2018, we deployed the final version to the online survey platform LimeSurvey. We used Prolific Academic to advertise the survey to UK residents only and at a rate of £7.50/h, to retrieve demographic information, and to reject answers below 10 minutes which we considered 'speeders'. The average time to finish was 15 minutes.

We collected information on our respondents' cohabitants (age, sex, occupation, relationship with the respondent) and which smart home devices the household owned. We then gauged attitudes toward internet-connected technology by asking for agreement with value-laden statements from our interviews \autocite{Kraemer2019}. The main survey had three parts: 
(A) unpacking the social context of roles and responsibilities in setting up and maintaining devices through five vignettes (factors included type of technology, the social relationship, and physical features); 
(B) three problem scenarios on device usage; and 
(C) two scenarios to explore moral and normative aspects of disclosing the use of smart cameras and voice assistants. 
In this paper, we report on findings from part (A) -- for details see Sidebar \ref{sidebar:survey-outline}.

\hypertarget{results}{%
\section{Results}\label{results}}

\marginpar{%

    {%
        \begin{minipage}{0.975\marginparwidth}
            \textbf{Attitude statements}\footnotesize
            \begin{flushleft}
                        \begin{enumerate}[leftmargin=15pt,label=(\alph*),itemsep=0pt,topsep=5pt]
            	
            	\item   Tech is my hobby. I will buy and try ... %new devices whenever I can. 
            	\item   I want to understand how these devices work, ... %what their benefits and limitations are. 
            	\item   Internet-connected devices are for younger people ...% who are more interested and engaged. 
            	\item   I like to do things myself and don’t need ...% technology to support me. 
            	\item   I can’t be bothered to find workarounds ... %if something doesn’t work as expected. 
            	\item   I have my reservation because of risks ... %attached to using technology. 
            	\item   If I had a choice, I’d rather not use any ... %technology like Wifi, computers and so on. 
            	\item       I’m happy to use ... [if they] improve my life % internet-connected (smart) devices as long as they improve my life. 
            \end{enumerate}
            \end{flushleft}

        \includegraphics[width=\linewidth,]{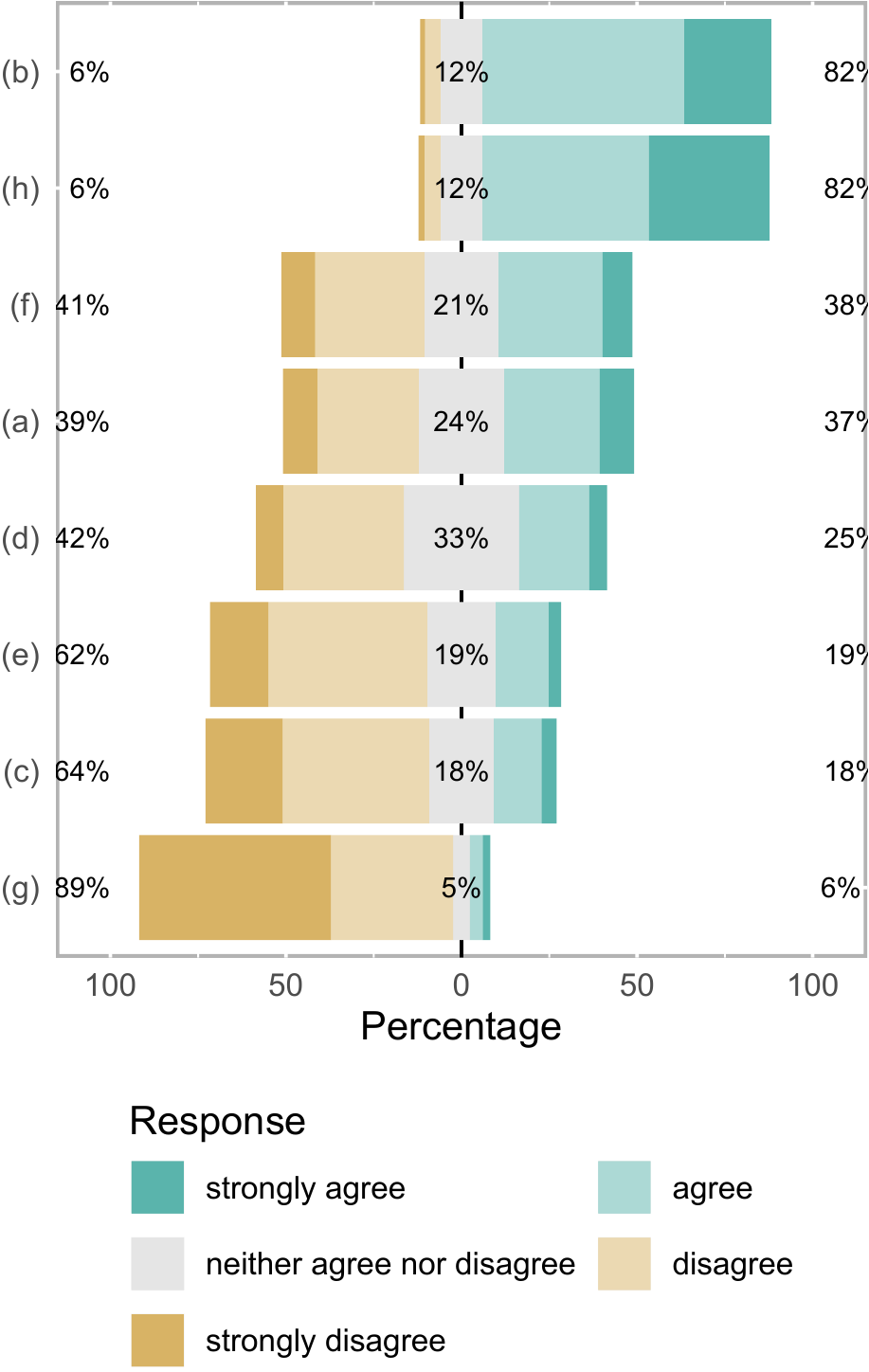} \captionof{figure}{Agreement with attitudes among respondents}\label{fig:people-attitudes}
        \end{minipage}
    }
}

In the final data set, 40\% of our respondents were 18-36 years old, 57\% were 37-64 years, and 3\% were above the age of 64; 52\% of respondents identified as male and 48\% as female. Our 850 respondents shared their homes with 1142 cohabitants (\textit{Mdn}=2, n=409). 48.8\% of households owned at least one device out of voice assistant, security systems, lights, sockets, thermostat, or vacuum cleaner.

\hypertarget{diverse-attitudes-in-households}{%
\subsection{Diverse attitudes in households}\label{diverse-attitudes-in-households}}

%While our respondents mostly maintained a positive attitude towards internet-connected devices, they reported differences with their cohabitants.
There were difference in attitude between respondents and their cohabitants.
82.1\% of our respondents agreed with (b) being interested in understanding benefits and limitations, and 81.7\% were (h) willing to use con\-nec\-ted technology to improve their lives (Figure \ref{fig:people-attitudes}). They did not want to use ``wifi, computers, and so on'' (89.3\%). Respondents were split on con\-nec\-ted technology being their hobby (38.7\% disagreed) and on whether they had reservations because of related risks (40.8\% disagreed).

%We also asked our respondents for the single most fitting statement to describe the attitudes in their households. 
We asked our respondents to describe their own and their cohabitants attitude by choosing the most appropriate statement (same as in Figure \ref{fig:people-attitudes}). To illustrate household complexity, we split the statements by sentiment into two groups:
%We split these statements by their sentiment in two groups to highlight the complexity of household attitudes: 
we consider (a), (b), and (h) to reflect interest and willingness to use technology (positive attitude) while (c), (d), (e), (f), and (g) represent caution and reluctance (negative attitude). About 76\% of all respondents were open or maintained a positive attitude (b: 27\%/h: 24\%/a: 25\%) while the remaining 24\% expressed some reservations. The respondents considered their cohabitants less positive about devices (a/b/f: 53\%) but more reserved (c/d/e/g/h: 47\%). Whithin households attitudes were diverse: 63.08\% of the 2-person households (n=409) shared a positive attitude; and among 3-person households (n=168), in 80.95\% at least two inhabitants shared a positive attitude; all three inhabitants shared a positive attitude in 36.31\%.

\hypertarget{attending-new-systems}{%
\subsection{Attending new systems}\label{attending-new-systems}}

These differences in attitude were somewhat reflected in administration responsibilities and perceived ability to use internet-connected devices. Respondents' considerations included the nature of the system itself, the ability, and the aptitude of individuals. In fact, often {more than one inhabitant was considered able to set up the system}: ``I would want to be responsible for setting up the system. Although my son would be equally capable and wanting to do it.'' (R716).
The majority (62.7\%) of our respondents would set up an internet-connected system (lights, thermostat, or security system) themselves or would trust close family members with the setup (25.5\%).

Among those setting up systems were 40.5\% male and 35-64 years old, of which 63\% were respondents and 33\% cohabiting family members. The majority of other cohabitants was considered willing to attempt the configuration (25.7\% equally able and willing / 38.5\% willing but needed help). Most of these people were female and immediate family members; and younger people were generally perceived as more able and willing than middle aged cohabitants.

\hypertarget{usage-expectations}{%
\subsection{Usage expectations}\label{usage-expectations}}

Our respondents expected most of their cohabitants to make use of an internet-connected system. Comments like ``the more i use the system the better; i and others would get use to it'' (R793) revealed thoughts on administrators' role and ability in helping others getting used to a system; and indeed most {inhabitants would use a system} (85.6\%). The majority of 70.0\% was considered able to use a system independently (15.7\% with help only), and the majority of those considered not interested in configuring a system in the first place (31.5\%), was anticipated to use the system later on (60.2\%).

\marginpar{%
	\vspace{-2cm}
	{%
		\begin{minipage}{\marginparwidth} 
			\textbf{Gender influence} \vspace{5pt} 

			{\footnotesize\centering Would anyone else be able to configure the system?  \\\vspace{5pt}
			\includegraphics[width=0.7\linewidth,trim=0cm 1.5cm 0 0,clip]{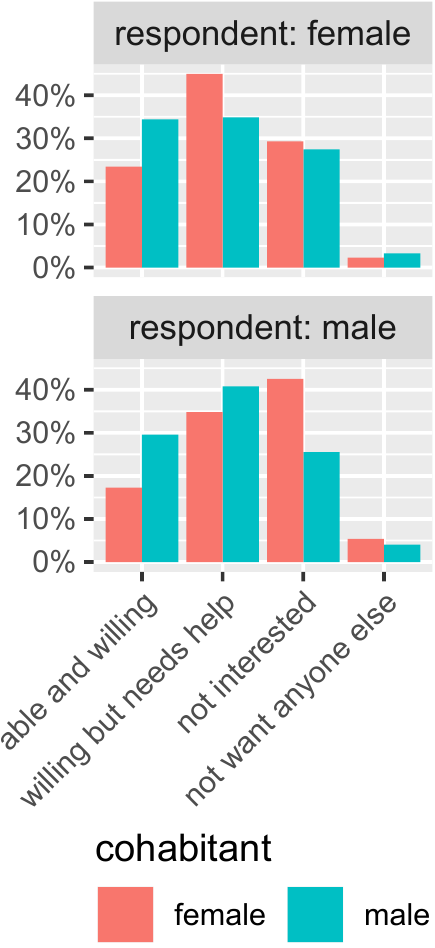} \\

			\vspace{10pt}How is the system going to be used? \\\vspace{5pt}
			\includegraphics[width=0.8\linewidth,]{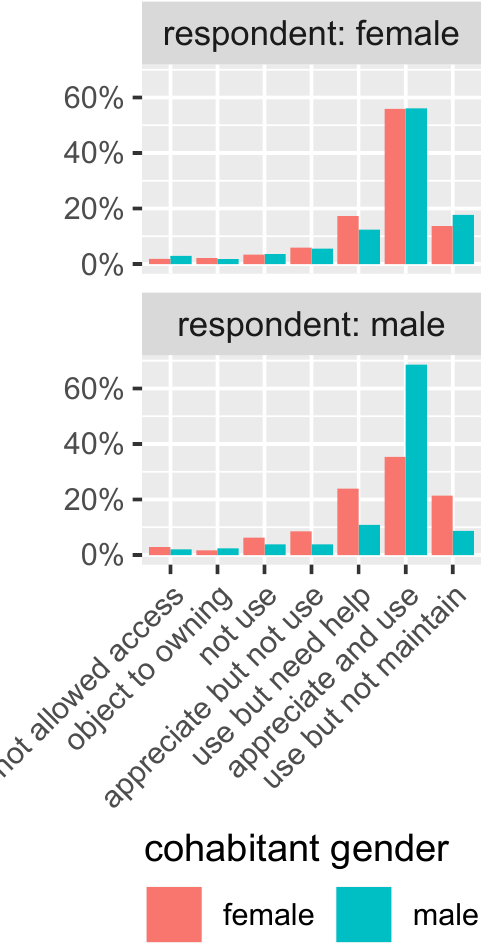} \captionof{figure}{Results of question (A-i): Differences in judging the same and opposite (colour codes) gender's ability to configure (top) and use (bottom) a system}\label{fig:gender-effects}}
		\end{minipage}
	} 
}

A closer look at questions (A.i.b+c) suggested this expectation was influenced by the respondent's and the cohabitant's gender (Figure \ref{fig:gender-effects}). Females were considered less able and willing to configure a system by male than by female respondents. In turn male respondents considered their male cohabitants more likely to be self-sufficient in using the system than their female cohabitants, while female respondents considered their male and female cohabitants equally self sufficient.
According to our female respondents (24.12\%), female non-administrators were more capable and willing than according to our male respondents (17.24\%); and female non-administrators were more likely to be willing but needing help in the eyes of our female respondents (44.73\%) than in the eyes of our male respondents (35.64\%). According to our male respondents, female non-administrators were more likely to be not interested (41.95\%) than according to our female respondents (28.95\%).

\hypertarget{relationships-and-hierarchies}{%
\subsection{Relationships and hierarchies}\label{relationships-and-hierarchies}}

\begin{figure}[t]
\centering
\includegraphics[width=\linewidth,]{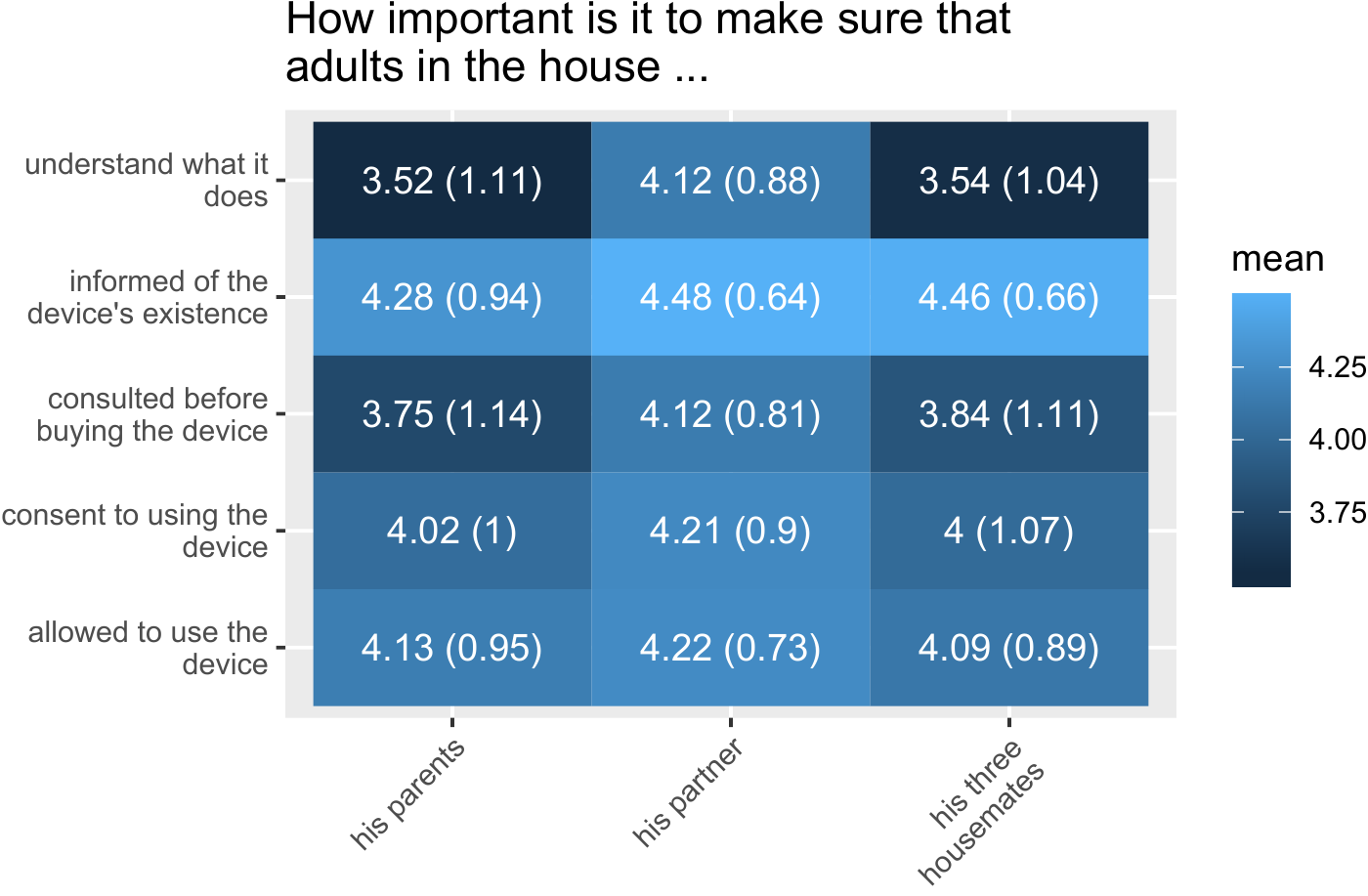} 
\caption{Mean agreement (standard deviation) on appropriate steps to introduce voice assistant when living with parents, partner, or three housemates. Participants agreed more on informing everyone than on consulting cohabitants and making sure they understand how the device works} \label{fig:30SN-overview}
\end{figure}

While most respondents expected their cohabitants to go along with the use of new devices, they attuned their responsibility to existing relationships and hierarchies by considering {trust, respect, ownership, and access}. They generally valued openness and transparency about intentions to purchase and use a particular device across all 15 vignettes (question (A.v)), with high mean scores for all answer options (Figure \ref{fig:30SN-overview} for smart voice assistants).

More specific relationship deliberations included: \emph{Partners} should generally consult each other on matters of new devices while sharing access to any existing device; and all items (see Figure \ref{fig:30SN-overview}) ranked high (\textgreater75\% agreement) across all technologies with almost no significant differences. Respondents presented with the \emph{housemates} scenario challenged whether devices were shared by emphasizing the importance of location (shared or private area). Respondents commented, regardless of their location, it was ``courtesy to consult'' (R817) flatmates on issues they could be affected by to avoid ``violations of trust'' (R835); housemates shared ``responsibilities and costs'' (R61). Particularly voice assistants and television required consideration of location as an indication to whether these devices were shared or not. There seemed to be uncertainty whether and to what extent the smart voice assistant in the scenario was a shared device (54.3\% agreed and 33.3\% were uncertain whether to allow use).
These considerations were amplified in the \emph{parents} scenario with stronger expressions of preference.
Respondents expected individuals to involve their cohabiting {parents} closely in the procurement of any security, light, or thermostat system (\textgreater75\% agreement for all items). In case of security and light systems, it was impossible to operate them without parents' approval. Strong comments surfaced in relation to voice assistants and referred to discomfort with the devices' presence due to privacy concerns, in one comment explicitly linked to the respondents' responsibility as a parent. The comments and scores suggested television (63.3\% allowed use) and voice assistant (51.3\% allowed use) could also be personal devices, and parents did not need to be allowed access.

\marginpar{%
	{%
		\begin{minipage}{0.975\marginparwidth} 
			\textbf{Social relationships} \vspace{5pt} 
			
			{\footnotesize\centering 
			\includegraphics[width=\linewidth,]{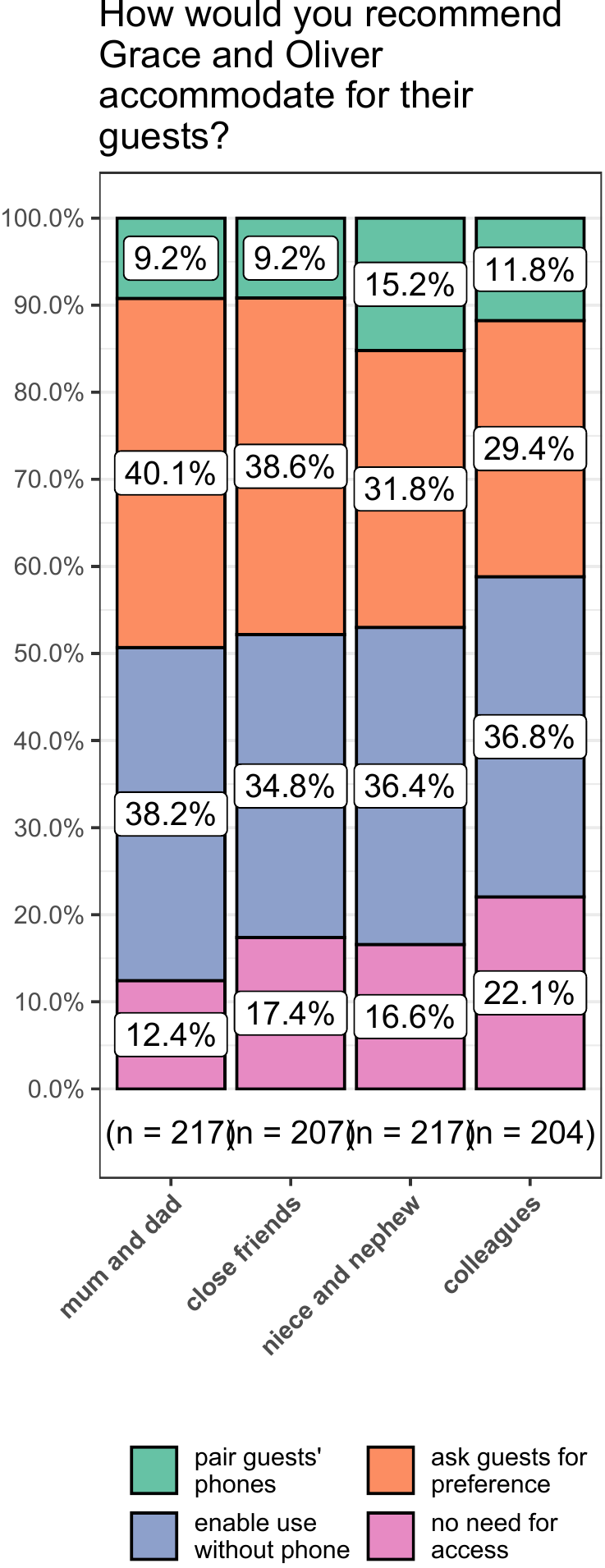} \captionof{figure}{Recommended adjustments for a smart home setup that relied on inhabitants' phones for control and information.}\label{fig:hospitality-consideration}}
		\end{minipage}
	}
}

\hypertarget{balancing-needs-and-demands}{%
\subsection{Balancing needs and demands}\label{balancing-needs-and-demands}}

Respondents considered trust and respect---explicitly or implicitly through established close relationships---as qualities of relationships; they referred to ability and skills when sharing access to internet-connected devices with their guests. Some explicitly highlighted the need to balance obligations of hospitality with their own security and privacy needs by considering responsibility and trustworthiness in their guests (Figure \ref{fig:hospitality-consideration}).

In this scenario, some features of a smart home system relied on inhabitants' phones for control and information. The owners considered adjustments to accommodate for their guests.
Respondents were most likely to ``strike the balance'' by enabling the system for use without a phone (36.57\%), followed by asking guests for their preference (35\%); 17\% thought there was no need for access, and 11\% agreed to pair their guests' phones.
Respondents preferred colleagues not to have access to their systems (58.9\%). The comments showed different attitudes ranging from disapproval of having to make any adjustments (``none of this should be necessary'' (R734)) to considerations of tech-savviness and system security (``giving access might compromise security'' (R436)). As relatives, \emph{nieces and nephews}, ``shouldn't be held responsible'' (R315) or be burdened with the system. While the younger relatives were considered tech-savvy enough to handle the system, respondents expressed concerns with regards to responsible use: They were not old enough to be trusted (R359). Respondents suggested to implement guest features for longer stays so that the system could be adjusted while security was maintained. While more likely to pair their phones (15.2\%) or ask for their preferences (31.8\%), most respondents preferred adjusting the system (36.4\%). \emph{Close friends} should not be bothered, and a balance between being polite and needs of security had to be found (``educate them and adjust'' (R809)). One respondent commented that ``asking for preference [was] mindful of guests privacy'' (R637). \emph{Parents}' needs were more important than the inhabitants', and the preferred way to accommodate them depended on what they were comfortable with. Some of the recommendations considered tech savviness of guests but also a need to keep parts of the home secure and private. 40.1\% recommended to ask the parents for their preference, closely followed by 38.2\% recommending to enable the system to be used without a phone.

\hypertarget{discussion}{%
\section{Discussion}\label{discussion}}

In our prior work \autocite{Kraemer2019}, we illustrated how household members considered qualities of devices, personal characteristics, and social groups in navigating technology use at home. The results of our survey show communal devices are unlikely to be used in a similar way of sameness, sharing them does not mean cohabitants have the same access; they show how aforementioned considerations are attuned to existing relationships and hierarchies with complex dynamics -- as common in social relationships of all kinds -- that can challenge intentions and assumptions underpinning the design of technology.

Firstly, our results highlight how technology related considerations fall in line with dominating social obligations, i.e.~those of parenting or being a good host. Comments on the use of smart voice assistants at home were remarkably strong in the parent-child vignettes and with regards to security and privacy aspects. The convention to be a good host and making guests comfortable dominated considerations of security and privacy. Striking a balance between being polite and maintaining personal levels of security and privacy was desired; for those less familiar with such technical issue, the relationship type might even be the only deciding factor in modelling use of communal devices for guests.

Secondly, these obligations and expectations are perceived as more clearly defined in established contexts, e.g.~parents and their children. Similar to \cite{Garg2019}, we find the more loosely defined relationships require more complex considerations to gauge qualities of relationships such as trust and reliability. More flexible configurations require more effort and can lead to dissatisfaction among system owners and administrators (``none of this should be necessary'' (R734)). On this basis, we suggest designing with social expectations in mind (e.g.~parents' needs are probably more important than yours (cf. R698)) can help guide design features that help navigate complex social situations. However, our findings also suggest the importance of communication and education, particularly in highlighting differences in perception of abilities and attitudes.

\marginpar{
\vspace{-1cm}
\small\textbf{Acknowledgements}\\
The data collection was funded by a Research Institute in Sociotechnical Cyber Security (RISCS) small grant. The first author was supported by an EPSRC grant (EP/P00881X/1).}

Thirdly, these expectations need to match assumptions of individual or communal use which in turn are influenced by social considerations and obligations. Such considerations are particularly important if the device in question is inherently designed for communal use since householders share ``cost and responsibility'' in using them. Voice assistants and television can be seen as personal or shared devices, and people consider factors like location as proxy to understand aforementioned dimensions of use, e.g.~a voice assistant in a personal bedroom is likely personal. Many devices already ask for their location and this information seems could be used to improve interaction.

\hypertarget{limitations}{%
\subsection{Limitations}\label{limitations}}

Firstly, we studied technology use with an online survey and by recruiting from an online platform. Our results confirm that our respondents are rather technology astute.
Secondly, our results are exploratory, and confirmatory follow-up studies would be needed to assess the replicability of specific findings.
Thirdly, we excluded `prefer not to say' as answer option from the presented results, leading to smaller sample sizes.

\hypertarget{conclusion}{%
\section{Conclusion and Future Work}\label{conclusion}}

Our results illustrate how social and technology considerations can influence expectations and intentions of technology use. They demonstrate how people apply social obligations and expectations in an well-intentioned effort to navigate communal use of devices. We suggest a design perspective of social expectations to structure loosely defined social relationship, and we highlight---like others before us---how aspects such as trust and responsibility or location can be used to anticipate degrees of communal and individual use.

We hope to publish a comprehensive analysis of the survey in the near future, and we will continue our investigation of communal use to empower privacy practices in the smart home~\cite{Kraemer2019a}.

\bibliography{bibliography}
\bibliographystyle{SIGCHI-Reference-Format}

\end{document}